\documentclass[fleqn,preprint,showpacs,preprintnumbers,amsmath,amssymb]{revtex4}
\usepackage{graphicx}
\usepackage{dcolumn}
\usepackage{bm}

\begin{document}

\title{Propagation of arbitrary amplitude nonlinear quantum ion-acoustic waves in electron-ion plasmas: Dimensionality effects}
\author{M. Akbari-Moghanjoughi}
\affiliation{Azarbaijan University of
Tarbiat Moallem, Faculty of Sciences,
Department of physics, 51745-406, Tabriz, Iran}

\date{\today}
\begin{abstract}
Propagation of arbitrary-amplitude ion-acoustic solitary (IASWs) as well as periodic waves (IAPWs) is investigated in a fully degenerate quantum electron-ion plasma consisting of isothermal- or adiabatic-ion species. It is shown that the system dimensionality and degrees of freedom play critical roles in matching criteria for propagation of such waves. Furthermore, it is revealed that for the case of adiabatic-ion unlike isothermal one, in some cases, there exists an upper fractional ion-temperature limit for the existence of IAPWs. It is also shown that, the variations of wave-amplitude with respect to the change in fractional ion-temperature is quite different for the cases of isothermal and adiabatic-ion plasmas.\\
\\
Keywords: Quantum plasma, Nonlinear Quantum ion-acoustic wave, Quantum hydrodynamics model, Sagdeev potential

\end{abstract}

\keywords{\textbf{Quantum plasma, Nonlinear Quantum ion-acoustic wave, Quantum hydrodynamics model, Sagdeev potential}}

\pacs{52.30.Ex, 52.35.-g, 52.35.Fp, 52.35.Mw}
\maketitle

\section{Introduction}

Ion acoustic waves (IAWs) have undeniable impact on scientific knowledge of characteristics of ionized environments. Linear and nonlinear features of these identities has been studied for long time since 1961 \cite{vedenov} using diverse techniques, of those Sagdeev \cite{popel, nejoh, mahmood1, mahmood2, mahmood3} and perturbation \cite{salah, Esfand2, Esfand3, Tiwari1, Tiwari2, Mushtaq} methods are well-known. Recent achievements in micro- and nano-electronic device fabrication has motivated new efforts to study the interesting nonlinear behavior of IAWs in dense plasma environments such as quantum dots, quantum wires, quantum wells, carbon nano-tubes, quantum diodes, ultra-cold plasmas, micro plasmas, biophotonic, etc. \cite{haug}. For instance, L. C. Gardner \cite{gardner} has applied quantum hydrodynamic (QHD) model to study the electron-hole dynamics in semiconductors.

Dense or degenerate warm plasmas are also abundant in astrophysical environments such as the core of planets, white dwarfs and neutron stars \cite{shapiro}. It has been shown that in a dense plasma, where the quantum mechanical rules are in action, new nonlinear effects arise \cite{haas, chatterjee, akbari} due to the non-classical quantum tunneling effect \cite{manfredi}. One of important parameters of a quantum plasma is the electron number-density which is a central parameter in definition of degenerate plasmas. As the number of electrons increase in quantum plasmas, quantum effects dominate due to Pauli exclusion principle leading to higher degeneracy pressure which is much larger than classical electron and ion pressures. Therefore, it is naturally expected that the system dimension and degrees of freedom to play crucial role in quantum plasma environments.

In recent few years there has been intense investigations on nonlinear behavior of planar and non-planar IAWs in quantum plasmas due to their experimental as well as theoretical importance. Recent investigations indicate that the propagation characteristics of solitary waves in bounded quantum plasma with non-planar geometry significantly differ from those in unbounded planar one \cite{misra, masood, sahu}. Characteristic features of linear and nonlinear propagation of IAWs has been studied extensively in electron-ion \cite{mahmood4, mahmood5} as well as electron-positron-ion quantum plasmas \cite{ali, masood2}. It has been found that only rarefactive large amplitude solitary excitations exist in quantum electron-ion plasmas, while, for electron-positron-ion plasmas only compressive large amplitude solitary excitations are allowed for given fractional plasma parameters.

On the other hand, it seems that less attention has been paid to the system dimensionality and \textbf{ion degrees of freedom effects}, which are supposed to be of vital importance in quantum systems. In current research we aim at exploring these effects on propagation of large-amplitude IAWs in an un-magnetized electron-ion quantum plasma. This investigation can be of sufficient importance in laboratory plasma research. The organization of the article is as follows. The basic normalized plasma equations are introduced in section \ref{equations}. Nonlinear arbitrary-amplitude solutions are derived in section \ref{Sagdeev}. Numerical analysis and discussions is given in section \ref{discussion} and final remarks are presented in section \ref{conclusion}.

\section{The Basic Equations}\label{equations}

Lets consider a dense plasma consisting of \textbf{inertialless} degenerate electrons and inertial positive hot ions. It is assumed that the plasma is fully degenerate so that the ion temperature is much less than the characteristic Fermi-temperatures of electrons. It is also noted that in a fully degenerate plasma the rate of electron-ion collisions are limited due to Fermi blocking process, hence, the plasma is considered almost collision-less. Therefore, it is reasonable to use an isothermal pressure model for ions. Using the conventional fluid quantum hydrodynamics (QHD) basic equations, we have for both electrons and ions
\begin{equation}\label{normal}
\begin{array}{l}
\frac{{\partial {n_i}}}{{\partial t}} + \frac{{\partial {n_i}{v_i}}}{{\partial x}} = 0, \\
\frac{{\partial {n_e}}}{{\partial t}} + \frac{{\partial {n_e}{v_e}}}{{\partial x}} = 0, \\
\frac{{\partial {v_i}}}{{\partial t}} + {v_i}\frac{{\partial {v_i}}}{{\partial x}} =  - \frac{e}{{{m_i}}}\frac{{\partial \phi }}{{\partial x}} - \left( {\frac{{\gamma{k_B}{T_i}}}{{{m_i}{n^{(0)}}}}} \right) \left(\frac{{{n_i}}}{n^{(0)}}\right )^{\gamma  - 2}\frac{{\partial {n_i}}}{{\partial x}} + \frac{{{\hbar ^2}}}{{2{m_i}^2}}\frac{\partial }{{\partial x}}\left[ {\frac{{{\partial ^2}/\partial {x^2}\left[ {\sqrt {{n_i}} } \right]}}{{\sqrt {{n_i}} }}} \right],\\
\frac{{{m_e}}}{{{m_i}}}\left[ {\frac{{\partial {v_e}}}{{\partial t}} + {v_e}\frac{{\partial {v_e}}}{{\partial x}}} \right] = 0 = \frac{e}{{{m_i}}}\frac{{\partial \phi }}{{\partial x}} - \frac{1}{{{m_i n_e}}}\frac{{\partial {P_e}}}{{\partial x}} + \frac{{{\hbar ^2}}}{{2{m_e}{m_i}}}\frac{\partial }{{\partial x}}\left[ {\frac{{{\partial ^2}/\partial {x^2}\left[ {\sqrt {{n_e}} } \right]}}{{\sqrt {{n_e}} }}} \right], \\
\frac{{{\partial ^2}\phi }}{{\partial {x^2}}} = 4\pi e\left( {{n_e} - {n_i}} \right), \\
\end{array}
\end{equation}
where $\gamma=1$ ($\gamma\neq 1$) with $\gamma=(f+2)/f$ ($f$ being the ion degrees of freedom) corresponds to isothermal (adiabatic) ions case. Note that the left hand-side of momentum equation for electrons has been set to zero due to \textbf{small electron-to-ion mass} ratio. By using the following scalings
\begin{equation}
x \to \frac{{{c_{s}}}}{{{\omega _{pi}}}}\bar x,\hspace{3mm}t \to \frac{{\bar t}}{{{\omega _{pi}}}},\hspace{3mm}n_{e,i} \to \bar n_{e,i}{n^{(0)}},\hspace{3mm}v \to \bar v{c_{s}},\hspace{3mm}\phi  \to \bar \phi \frac{{2{E_{Fe }}}}{e},
\end{equation}
\textbf{where, ${\omega _{pi }} = \sqrt {4\pi{e^2}n_{e}^{(0)}/{m_i}}$} and ${c_{s}} = \sqrt {2{E_{Fe }}/{m_i }}$ are characteristic plasma-frequency and quantum sound-speed ($E_{Fe}=k_B T_{Fe}$ is the electron Fermi-energy), respectively, we obtain the normalized basic QHD equation set as
\begin{equation}\label{normal}
\begin{array}{l}
\frac{{\partial {n_i}}}{{\partial t}} + \frac{{\partial {n_i}{v_i}}}{{\partial x}} = 0, \\
\frac{{\partial {n_e}}}{{\partial t}} + \frac{{\partial {n_e}{v_e}}}{{\partial x}} = 0, \\
\frac{{\partial {v_i}}}{{\partial t}} + {v_i}\frac{{\partial {v_i}}}{{\partial x}} =  - \frac{{\partial \phi }}{{\partial x}} - \sigma \frac{\partial }{{\partial x}}\ln {n_i}, (\gamma= 1)\\
\frac{{\partial {v_i}}}{{\partial t}} + {v_i}\frac{{\partial {v_i}}}{{\partial x}} =  - \frac{{\partial \phi }}{{\partial x}} - \left( {\frac{{\gamma \sigma }}{{\gamma  - 1}}} \right)\frac{{\partial {n_i}^{\gamma  - 1}}}{{\partial x}},(\gamma  \ne 1)\\
\frac{{\partial \phi }}{{\partial x}} - \frac{1}{{{2E_{Fe}n^{(0)} n_i}}}\frac{{\partial {P_e}}}{{\partial x}} + \frac{{{H^2}}}{2}\frac{\partial }{{\partial x}}\left[ {\frac{{{\partial ^2}/\partial {x^2}\left[ {\sqrt {{n_e}} } \right]}}{{\sqrt {{n_e}} }}} \right] = 0, \\
\frac{{{\partial ^2}\phi }}{{\partial {x^2}}} = {n_e} - {n_i}, \\
\end{array}
\end{equation}
\textbf{where the fractional temperature $\sigma$ ($\sigma  = \frac{{{T_i}}}{{2{T_{Fe}}}}$)} measures the ion temperature relative to the electron Fermi-temperature and is much less than unity in a fully-degenerate zero-temperature Fermi-gas. The new parameter, $H=\hbar \omega_{pe}/2E_{Fe}$, is called the quantum diffraction parameter which is the ratio of plasmon-energy to electron Fermi-energy. Note also that, we have ignored the quantum term (last-term) in momentum equation of inertial heavy ions.

One notices that, under the zero-temperature assumption in a dense and degenerate plasma environment, the degeneracy pressure follows from Pauli exclusion principle and relates to the particle number-density \textbf{through the following relation}
\begin{equation}\label{P}
{P_e } = \frac{{2E_{Fe }}{{n ^{(0)}}}}{{(d + 2)}}{{n_e }^{\left(\frac{{d + 2}}{d}\right)}},
\end{equation}
where, $d$ represents the dimensionality of the system. Introducing the new variable $\xi=x-Mt$ ($M$ is a measure of waves group-speed relative to that of quantum sound, $c_s$), and solving for the ion density and momentum equations altogether, we get
\begin{equation}\label{ni}
\begin{array}{l}
\phi  = \frac{{{M^2}}}{2}\left[ {1 - \frac{1}{{{n_i}^2}}} \right] - \sigma \ln {n_i},(\gamma= 1), \\
\phi  = \frac{{{M^2}}}{2}\left[ {1 - \frac{1}{{{n_i}^2}}} \right] - \frac{{\gamma \sigma }}{{\gamma  - 1}}\left[ {1 - {n_i}^{\gamma  - 1}} \right],(\gamma\neq 1).
\end{array}
\end{equation}
\textbf{Moreover,} using Eq. (\ref{P}), in the electrons \textbf{energy relation}, after integrating once with appropriate boundary conditions, we obtain
\begin{equation}\label{ne}
\phi  =  - \frac{1}{2} + \frac{{{n_e}^{\frac{2}{d}}}}{2} - \frac{{{H^2}}}{{2\sqrt {{n_e}} }}\frac{{{\partial ^2}\sqrt {{n_e}} }}{{\partial {\xi^2}}}.
\end{equation}
It is evident that the conditions $\mathop {\lim }\limits_{{v_i} \to 0} {n_{i,e}} = 1$ and $\mathop {\lim }\limits_{{v_i} \to 0} \phi  = 0$ are met.

\section{Arbitrary-amplitude Nonlinear Waves}\label{Sagdeev}

In this section we will employ Sagdeev approach to find a suitable pseudo-potential which appropriately describes the propagation of ion-acoustic (periodic and solitary) waves and obtain some criteria for existence of such waves. This is done by evaluation of Eqs. (\ref{ni}) and (\ref{ne}) at local near-equilibrium state $n_e \approx n_i=n$. Using the above mentioned equations and defining new variable $N=\sqrt{n}$ yields
\begin{equation}\label{equi}
\begin{array}{l}
\frac{{{\partial ^2}N}}{{\partial {\xi ^2}}} = \frac{2}{{{H^2}}}\left[ {\frac{{{N^{\left( {\frac{{d + 4}}{d}} \right)}}}}{2} - \frac{N}{2} - \frac{{{M^2}}}{2}\left( {N - {N^{ - 3}}} \right) + \sigma N\ln {N^2}} \right],(\gamma  = 1), \\
\frac{{{\partial ^2}N}}{{\partial {\xi ^2}}} = \frac{2}{{{H^2}}}\left[ {\frac{{{N^{\left( {\frac{{d + 4}}{d}} \right)}}}}{2} - \frac{N}{2} - \frac{{{M^2}}}{2}\left( {N - {N^{ - 3}}} \right) + \frac{{\gamma \sigma }}{{\gamma  - 1}}\left( {N - {N^{2\gamma  - 1}}} \right)} \right],(\gamma  \ne 1), \\
\end{array}
\end{equation}
or equivalently,
\begin{equation}
\begin{array}{l}
\frac{1}{2}{\left( {\frac{{\partial N}}{{\partial \xi }}} \right)^2} = \frac{2}{{{H^2}}}\int {\left[ {\frac{{{N^{\left( {\frac{{d + 4}}{d}} \right)}}}}{2} - \frac{N}{2} - \frac{{{M^2}}}{2}\left( {N - {N^{ - 3}}} \right) + \sigma N\ln {N^2}} \right]dN}  =  - U(N),(\gamma  = 1), \\
\frac{1}{2}{\left( {\frac{{\partial N}}{{\partial \xi }}} \right)^2} =\\ \frac{2}{{{H^2}}}\int {\left[ {\frac{{{N^{\left( {\frac{{d + 4}}{d}} \right)}}}}{2} - \frac{N}{2} - \frac{{{M^2}}}{2}\left( {N - {N^{ - 3}}} \right) + \frac{{\gamma \sigma }}{{\gamma  - 1}}\left( {N - {N^{2\gamma  - 1}}} \right)} \right]dN =  - U(N),(\gamma  \ne 1).}  \\
\end{array}
\end{equation}
The well-known energy integral reads as
\begin{equation}
\begin{array}{l}
\frac{1}{2}{\left( {\frac{{\partial n}}{{\partial \xi }}} \right)^2} + U(n) = 0, \\
U(n) =  - \left[ {\frac{{4n + 2d{n^{2\left( {\frac{{d + 1}}{d}} \right)}} - 2(2 + d)({M^2}{{(1 - n)}^2} + n(n - 2(1 - n)\sigma )) + 4(2 + d)\sigma {n^2}\ln (n)}}{{(2 + d){H^2}}}} \right],(\gamma= 1), \\
U(n) =  - \left[ {\frac{{2n(\gamma  - 1)(d{n^{\frac{{d + 2}}{d}}} - (2 + d)\sigma  + 1) - 2(2 + d)\left( {(\gamma  - 1){{(1 - n)}^2}{M^2} - 2\sigma {n^{\gamma  + 1}} + {n^2}\left( {1 - \gamma  + 2\gamma \sigma } \right)} \right)}}{{(2 + d)(\gamma  - 1){H^2}}}} \right],(\gamma\neq 1).
\end{array}
\end{equation}
It is clearly noticed that the potential and its first derivative vanish at $n=1$. The possibility of ion-acoustic solitary waves require the following conditions to simultaneously met
\begin{equation}\label{conditions}
{\left. {U(n)} \right|_{n = 1}} = {\left. {\frac{{dU(n)}}{{dn}}} \right|_{n = 1}} = 0,{\left. {\frac{{{d^2}U(n)}}{{d{n^2}}}} \right|_{n = 1}} < 0.
\end{equation}
It is further required that for at least one either maximum or minimum nonzero $n$-value, we have $U(n_{m})=0$, so that for every value of $n$ in the range ${n _m} > n  > 1$ or ${n _m} < n  < 1$, $U(n)$ is negative. In such a condition we can obtain a potential minimum which describes the possibility of an IASW propagation.

On the other hand, for existing a periodic IAW we require that
\begin{equation}\label{conditions2}
{\left. {U(n)} \right|_{n = 1}} = {\left. {\frac{{dU(n)}}{{dn}}} \right|_{n = 1}} = 0,{\left. {\frac{{{d^2}U(n)}}{{d{n^2}}}} \right|_{n = 1}} > 0,
\end{equation}
and that, for at least one either maximum or minimum nonzero $n$-value, we have $U(n_{m})=0$, so that for every value of $n$ in the range ${n _m} > n  > 1$ or ${n _m} < n  < 1$, $U(n)$ is positive. In this situation we obtain a potential maximum describing the possibility of an IAPW propagation. Inspection of the second-derivative of pseudo-potential shows that
\begin{equation}\label{conditions2}
\begin{array}{l}
\text{For } \gamma=1\left\{ {\begin{array}{*{20}{c}}
   {{d^2}U(n)/d{n^2}\left| {_{n = 1}} \right. < 0\begin{array}{*{20}{c}}
   {} & {if\begin{array}{*{20}{c}}
   {} & {{M^2} - \sigma  < {1/d}}  \\
\end{array}}  \\
\end{array}}  \\
   {{d^2}U(n)/d{n^2}\left| {_{n = 1}} \right. > 0\begin{array}{*{20}{c}}
   {} & {if\begin{array}{*{20}{c}}
   {} & {{M^2} - \sigma  > {1/d}}  \\
\end{array}}  \\
\end{array}}  \\
\end{array}} \right.\\
\text{For } \gamma\neq 1\left\{ {\begin{array}{*{20}{c}}
   {{d^2}U(n)/d{n^2}\left| {_{n = 1}} \right. < 0\begin{array}{*{20}{c}}
   {} & {if\begin{array}{*{20}{c}}
   {} & {{M^2} + \gamma\sigma < {1/d}}  \\
\end{array}}  \\
\end{array}}  \\
   {{d^2}U(n)/d{n^2}\left| {_{n = 1}} \right. > 0\begin{array}{*{20}{c}}
   {} & {if\begin{array}{*{20}{c}}
   {} & {{M^2} + \gamma\sigma > {1/d}}  \\
\end{array}}  \\
\end{array}}  \\
\end{array}} \right.
\end{array}
\end{equation}
It can be easily confirmed that for $\gamma=1$ we always have $\mathop {\lim }\limits_{n \to 0} U(n) = 2{M^2}/{H^2}>0$ and $\mathop {\lim }\limits_{n \to + \infty} U(n) = -\infty$ independent of other plasma parameters. On the other hand, for $\gamma\neq 1$ we have $\mathop {\lim }\limits_{n \to 0} U(n) = 2{M^2}/{H^2}>0$, independent of other plasma parameters and for $f>d$ we obtain $\mathop {\lim }\limits_{n \to + \infty} U(n) = -\infty$, but for $f=d$ we have $\mathop {\lim }\limits_{n \to + \infty} U(n) = -\infty$ if $\sigma<1/(d+2)$. It is therefore concluded that for IASW to exist we must have ${{M^2} - \sigma  < {1/d}}$ for $\gamma= 1$ and ${{M^2} + \gamma\sigma  < {1/d}}$ for $\gamma\neq 1$. On the other hand for the possibility of IAPWs for $\gamma=1$ we must have ${{M^2} - \sigma  > {1/d}}$ and in the case of $\gamma\neq 1$ with $f>d$ we must have ${{M^2} + \gamma\sigma  > {1/d}}$ and for $f=d$ we must have ${{M^2} + \gamma\sigma  > {1/d}}$ and also $\sigma<1/(d+2)$.

The stationary wave solutions corresponding to this pseudo-potential which satisfis the mentioned boundary-conditions, read as
\begin{equation}\label{soliton}
\xi  - {\xi _0} =  \pm \int_1^n {\frac{{dn}}{{\sqrt { - 2U(n)} }}}.
\end{equation}

\section{Numerical Analysis}\label{discussion}

Numerical evaluation of pseudo-potential profiles indicates that only rarefactive ($n_m<1$) large-amplitude solitary waves exist in this quantum electron-ion plasma for all system dimensionality ($d$) and the ion degrees of freedom ($f$). This is in agreement with the previously reported results in Ref. \cite{chat} for quantum electron-ion plasma with the special case of $\gamma=3$ in one dimension. In current numerical scheme we only plotted the figures for the special case of (point-like ions) $d=f$. In Figs. 1(a) ($\gamma=1$) and 2(a) ($\gamma\neq 1$) the regions in $M$-$\sigma$ plane is shown, where IASWs or IAPWs can occur. The curves in these figures are $M^2-\sigma=1/d$ and ${{M^2} + \gamma\sigma = {1/d}}$ borders for $\gamma=1$ and $\gamma\neq 1$, respectively, and the horizontal lines for $\gamma\neq 1$ define the upper $\sigma$-limits ($\sigma=1/(d+2)$) of IAPW existence for plasma with adiabatic point-like ($d=f$) ions. The left/right side of curves in Fig. 1(a) is the region where IASWs/IAPWs can exist. On the other hand, for $\gamma\neq 1$ IAPWs stability region (${{M^2} + \gamma\sigma > {1/d}}$), shown in Fig. 2(a), is also bounded by an upper sigma limit ($\sigma<1/(d+2)$) for point-like ions $d=f$. Note that the $\sigma=0$ (cold-ion) limits which match exactly for the cases of plasma with isothermal and adiabatic ions gives a maximum/minimum ($M_m=1/\sqrt{d}$) for adiabatic/isothermal-ion plasma case. It is also clearly observed that for $d=1$ in isothermal-ion case, contrary to that of adiabatic-ion, there is always a region (for all possible values of fractional ion-temperature, $\sigma$) of supersonic solitary excitations, while for the shown range of $\sigma$ only subsonic IASWs exist for $d=2$ and $d=3$. However, for the case of adiabatic-ion plasma all solitary excitations lie in the subsonic region, for all dimensionality values. It is further remarked that the increase in the value of fractional ion-temperature increases/decreases the possible Mach-number range for all values of dimensionality for isothermal/adiabatic-ion case. It is observed that, as the dimension of the system is increased the possible Mach-number range for IASWs decreases for all values of $\sigma$ in both isothermal-ion and adiabatic-ion cases.

Figures 1(b-d) and 2(b-d) show the variations of the Sagdeev pseudo-potential profiles with respect to one-by-one variations of different fractional plasma parameters. For instance, it is concluded that for both $\gamma=1$ and $\gamma\neq 1$ cases the shape of pseudo-potential is significantly affected by the change in the system dimensionality, quantum diffraction parameter and fractional ion-temperature for fixed other plasma parameters. Figures 3(a-c) and 4(a-c) presents the solitary wave profiles corresponding to the pseudo-potential profiles shown in Figs. 1(b-d) and 2(b-d), respectively. From Figs. 3(a) and 4(a) it is clearly remarked that for both $\gamma=1$ and $\gamma\neq 1$ cases the IASW-amplitude decreases as the system dimension is increased, where, other plasma parameters are fixed. On the other hand, it is remarked from Figs. 3(b) and 4(b) that the soliton amplitude is unaffected by increase in the value of quantum diffraction parameter $H$, for all $\gamma$-values, however, its width is increased. This is a commonly reported feature of solitary excitations for many dense plasma environments \cite{haas, chatterjee, akbari}. Moreover, it is concluded from Figs. 3(c) and 4(c) that increase in the value of fractional ion-temperature increases/decreases the ion-acoustic soliton amplitude for isothermal/adiabatic-ion case. Figure 3(d) depicts the solitary profiles for sub- and supersonic cases (with $d=1$ and $\gamma=1$) which indicates that the supersonic soliton is shorter and wider compared to the subsonic counterpart. Also, Fig. 4(d) depicts the solitary profiles for isothermal (thick curve) and adiabatic (thin curve) cases (with similar other parameters) which indicates that the isothermal-ion solitary wave is slightly higher and narrower compared to that of adiabatic-ion.

Figures 5(a) and 6(a) reveal that the increase in the Mach-number in the case of periodic ion-acoustic waves causes the wave amplitude to increase. A comparison has been made between pseudo-potential profiles in Figs. 1(b-d)/2(b-d) for IASWs and those in Figs. 5(b-d)/6(b-d) for IAPWs, respectively. In addition to the large variation in pseudo-potential shape due to change in the system dimensionality for the case of IAPWs (compared to those of IASWs), it is observed that contrary to the case of IASWs the periodic ion-acoustic wave-amplitudes increase as the dimensionality increases. It is observed that the amplitude of IAPW and IASW are affected in a quite similar manner by the variations in the value of quantum diffraction parameter, $H$. A striking difference between IAPW and IASW for both $\gamma=1$ and $\gamma\neq 1$ cases is the behavior of amplitude variation due to change in the value of fractional ion-temperature, which can be noticed by comparing Figs. 1(d)/2(d) with 5(d)/6(d), respectively. It is evident that increase in the fractional ion-temperature increases/decreases the amplitude of IASWs for isothermal/adiabatic-ion case, while it decreases/increases the amplitude of IAPWs for isothermal/adiabatic-ion case.

\section{Conclusion}\label{conclusion}

Using the Sagdeev potential approach, we investigated the large-amplitude ion-acoustic solitary as well as periodic wave propagations in a degenerate quantum plasma with isothermal or adiabatic ion species in different plasma system dimensionality, taking into account the effect of ion-temperature and have shown that the system dimensionality and ion degrees of freedom has crucial impact on the propagation of such waves. In the studied range of fractional ion-temperature the one dimensional system with isothermal ions contains supersonic solitary propagations, while 2- and 3-D systems do not. For plasma with adiabatic ions only subsonic IASWs could be found for all dimensionality and for the chosen range of fractional ion-temperature. For the case of adiabatic-ion unlike isothermal one there is an additional upper fractional ion-temperature limit for the existence of IAPWs. It was also shown that, the variations of wave-amplitude with respect to the change in fractional ion-temperature is quite opposite for the cases of IASWs and IAPWs and for isothermal-ion and adiabatic-ion cases.

\newpage

\textbf{FIGURE CAPTIONS}

\bigskip

Figure-1

\bigskip

(Color online) (a) The stability regions of arbitrary-amplitude IASWs and IAPWs is shown in $M$-$\sigma$ plane with degenerate electrons and isothermal ions. The region to the left/right of curves indicate the stability region for IASW/IAPW. The thickness of the curve increases as the dimension becomes higher. (b), (c), (d) Variations in Sagdeev pseudo-potential profiles of IASW with respect to different plasma parameters. The change in the dash-size appropriately represents the change in the varied parameters in each plot.

\bigskip

Figure-2

\bigskip

(Color online) (a) The stability regions of arbitrary-amplitude IASWs and IAPWs is shown in $M$-$\sigma$ plane with degenerate electrons and adiabatic ions. The region to the below/above of curves indicate the stability region for IASW/IAPW. The horizontal lines correspond to the upper $\sigma$-limits of IAWs for the case of $d=f$ (point-like ions). The thickness of the curve increases as the dimension becomes higher. (b), (c), (d) Variations in Sagdeev pseudo-potential profiles of IASW with respect to different plasma parameters. The change in the dash-size appropriately represents the change in the varied parameters in each plot.

\bigskip

Figure-3

\bigskip

(Color online) The variation of the solitary wave profiles for isothermal-ion with respect to plasma parameters represented in Figs. 1(a-c). (d) Wave profiles for supersonic (thick) and subsonic (thin) solitary waves. The change in the curve-thickness appropriately represents the change in the varied parameters in each plot.

\bigskip

Figure-4

\bigskip

(Color online) The variation of the solitary wave profiles for adiabatic-ion case with respect to plasma parameters represented in Figs. 1(a-c). (d) Wave profiles for isothermal (thick) and adiabatic (thin) solitary waves with similar other plasma parameters. The change in the curve-thickness appropriately represents the change in the varied parameters in each plot.

\bigskip

Figure-5

\bigskip

(Color online) Variations in Sagdeev pseudo-potential profiles of IAPW for isothermal-ion case with respect to different plasma parameters. The change in the dash-size appropriately represents the change in the varied parameters in each plot.

\bigskip

Figure-6

\bigskip

(Color online) Variations in Sagdeev pseudo-potential profiles of IAPW for adiabatic-ion case with respect to different plasma parameters. The change in the dash-size appropriately represents the change in the varied parameters in each plot.

\newpage

\begin{figure}
\resizebox{1\columnwidth}{!}{\includegraphics{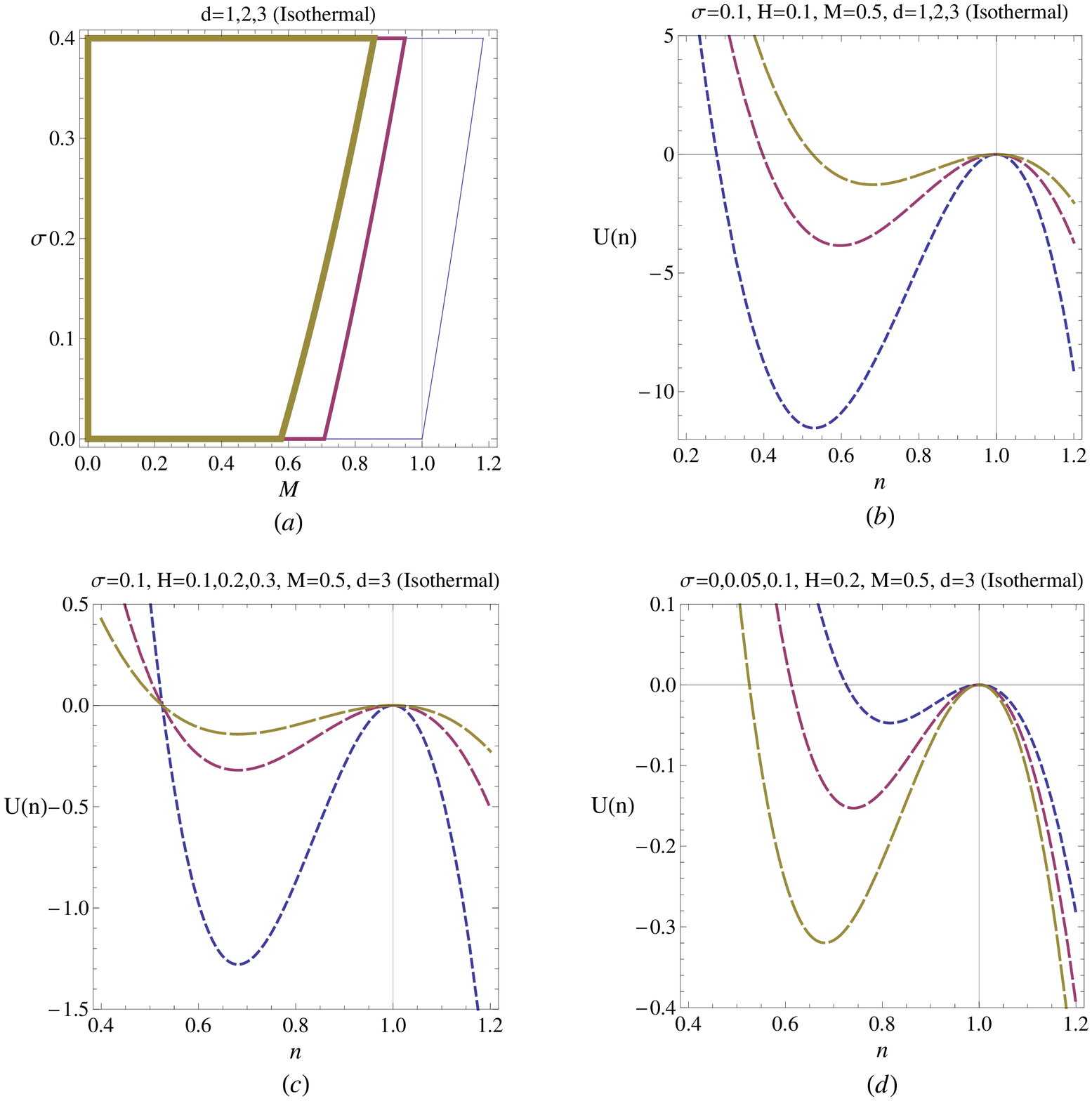}}
\caption{}
\label{fig:1}
\end{figure}

\newpage

\begin{figure}
\resizebox{1\columnwidth}{!}{\includegraphics{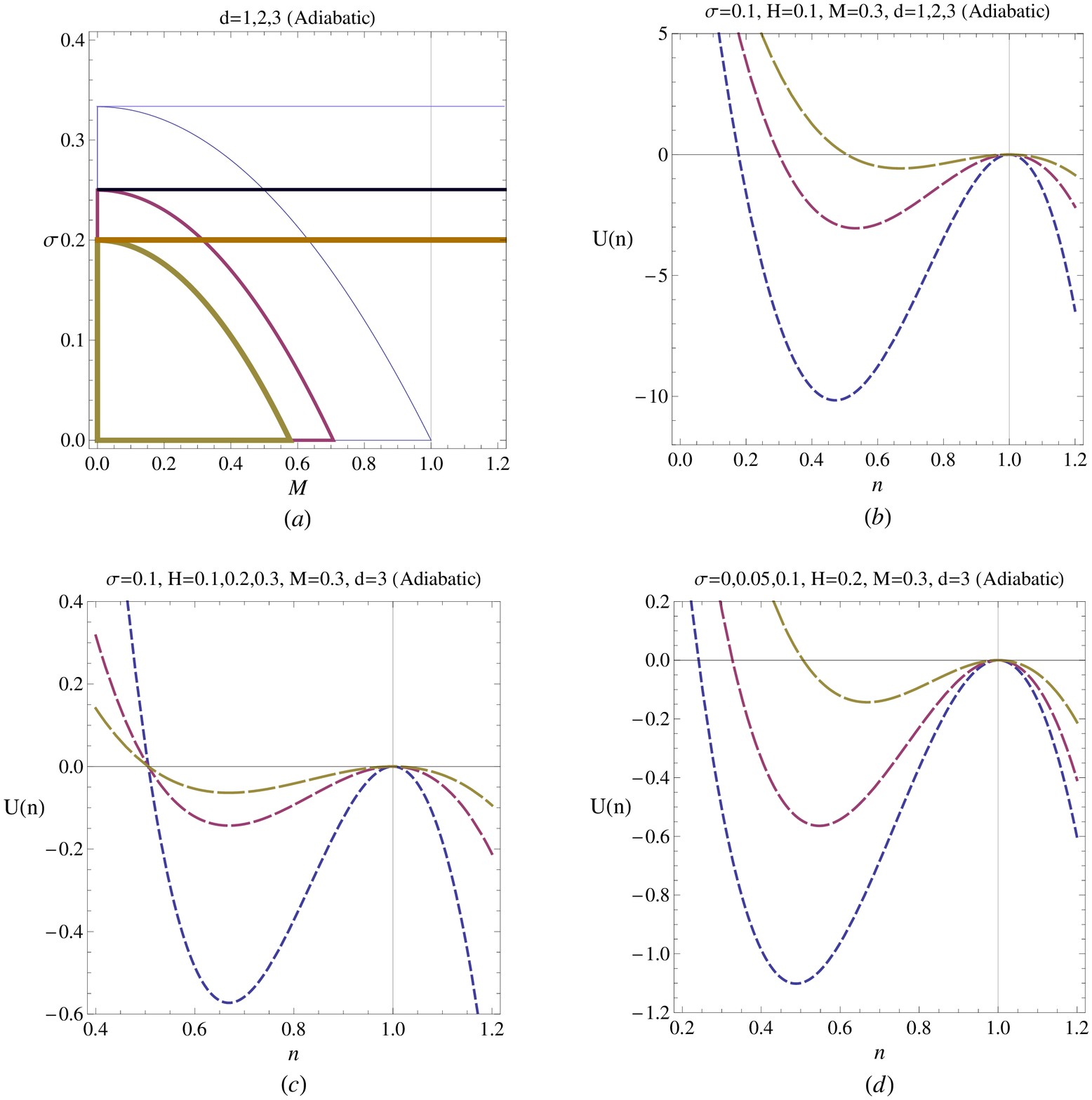}}
\caption{}
\label{fig:2}
\end{figure}

\newpage

\begin{figure}
\resizebox{1\columnwidth}{!}{\includegraphics{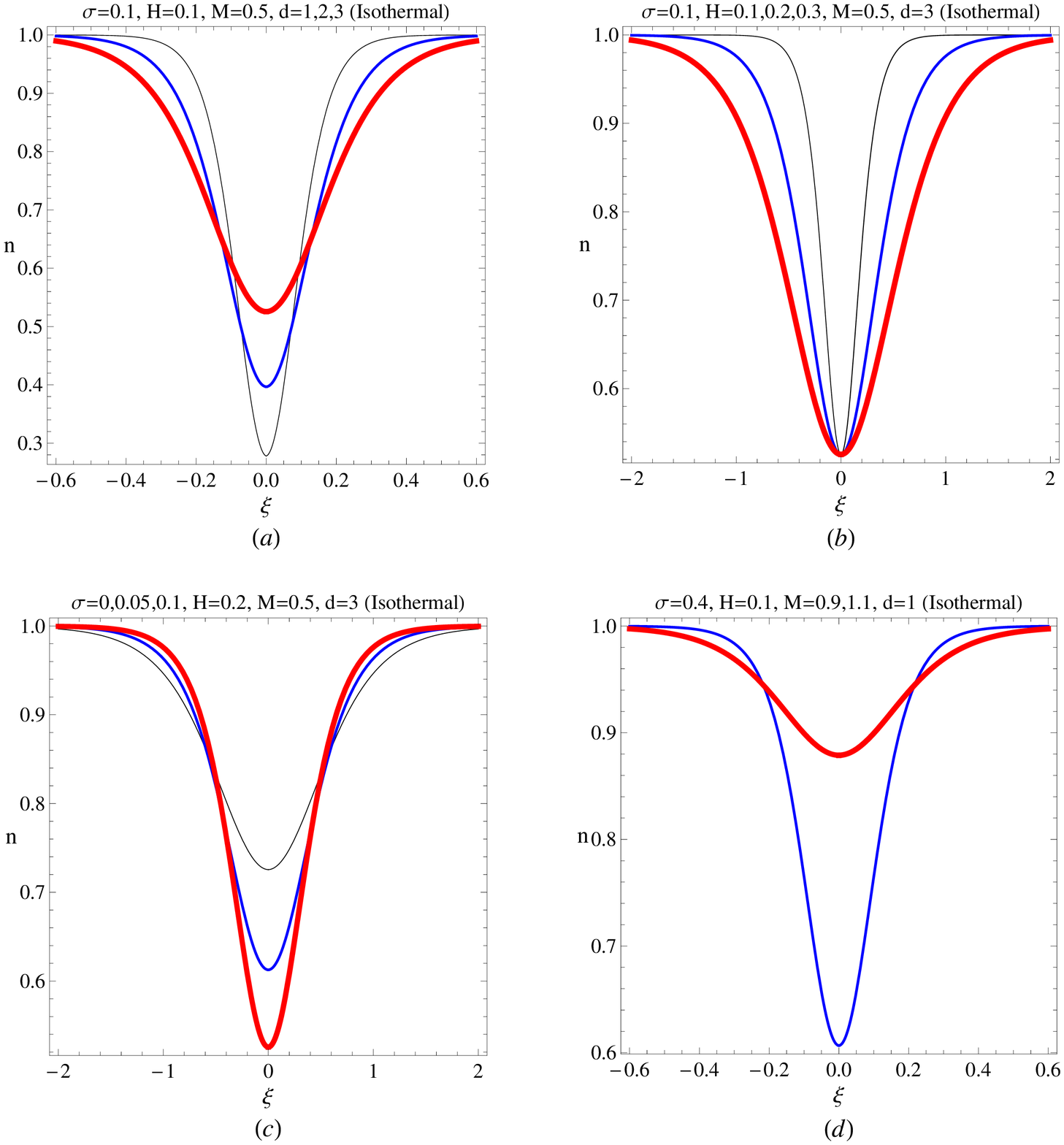}}
\caption{}
\label{fig:3}
\end{figure}

\newpage

\begin{figure}
\resizebox{1\columnwidth}{!}{\includegraphics{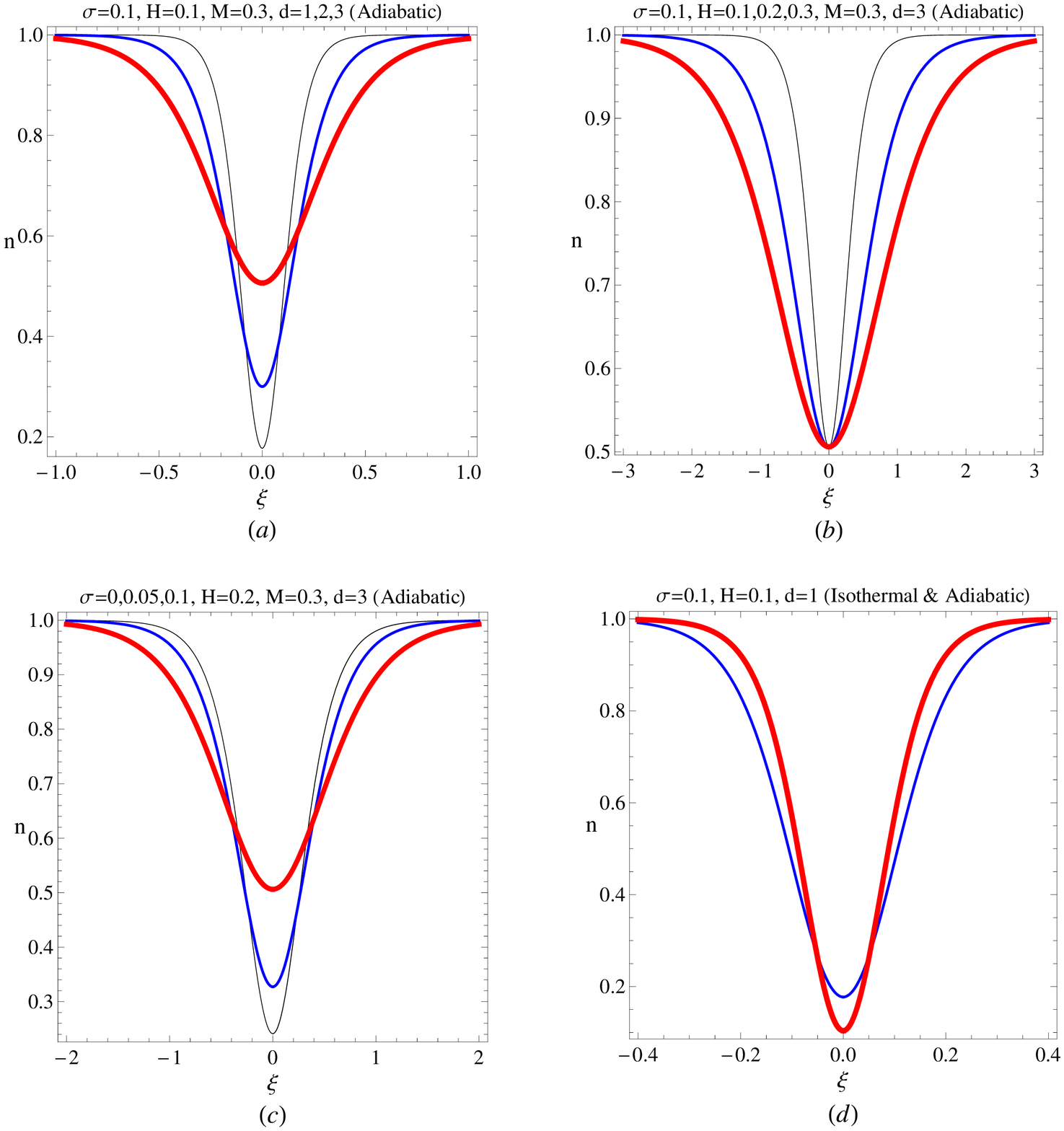}}
\caption{}
\label{fig:4}
\end{figure}

\newpage

\begin{figure}
\resizebox{1\columnwidth}{!}{\includegraphics{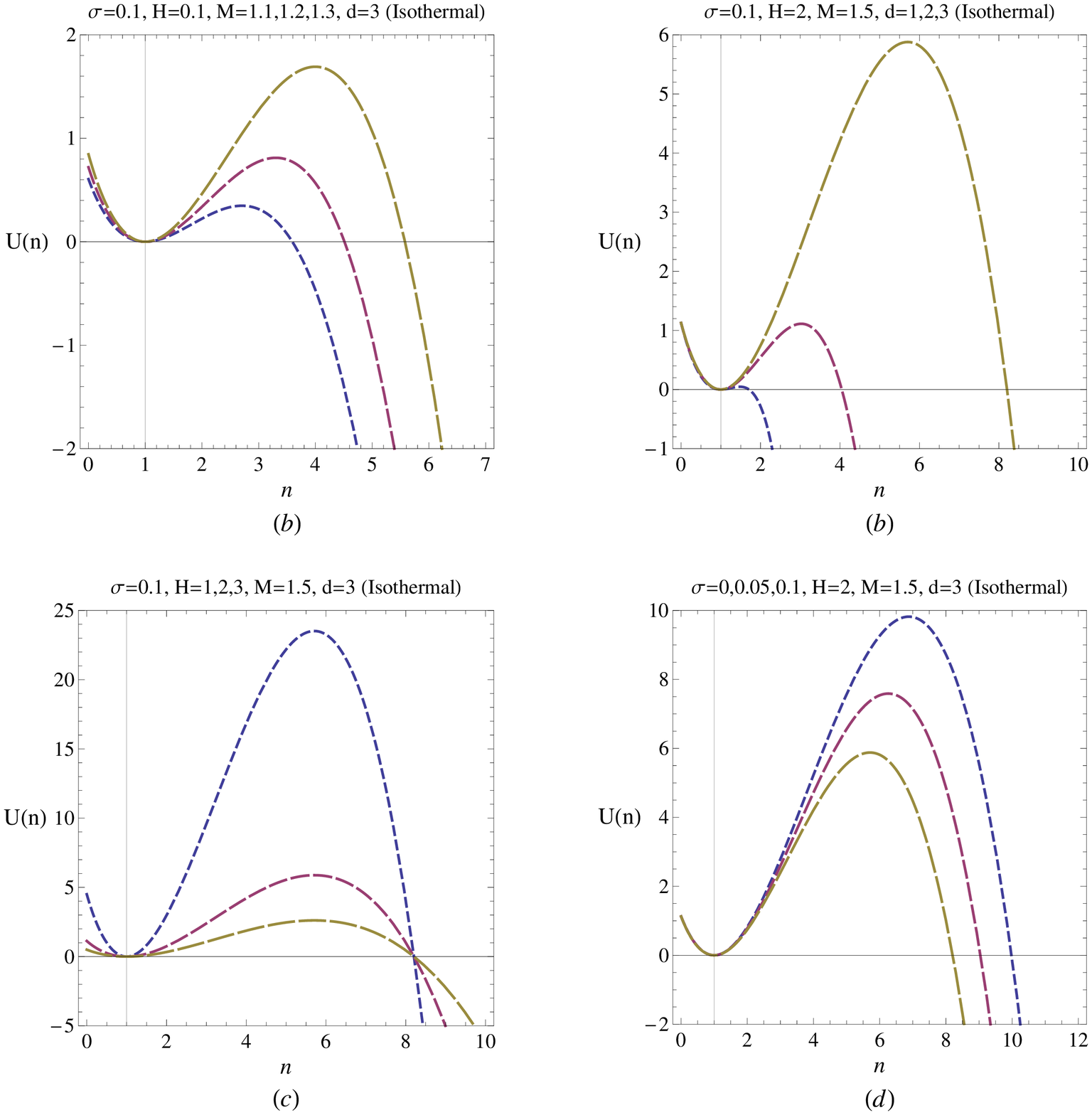}}
\caption{}
\label{fig:5}
\end{figure}

\newpage

\begin{figure}
\resizebox{1\columnwidth}{!}{\includegraphics{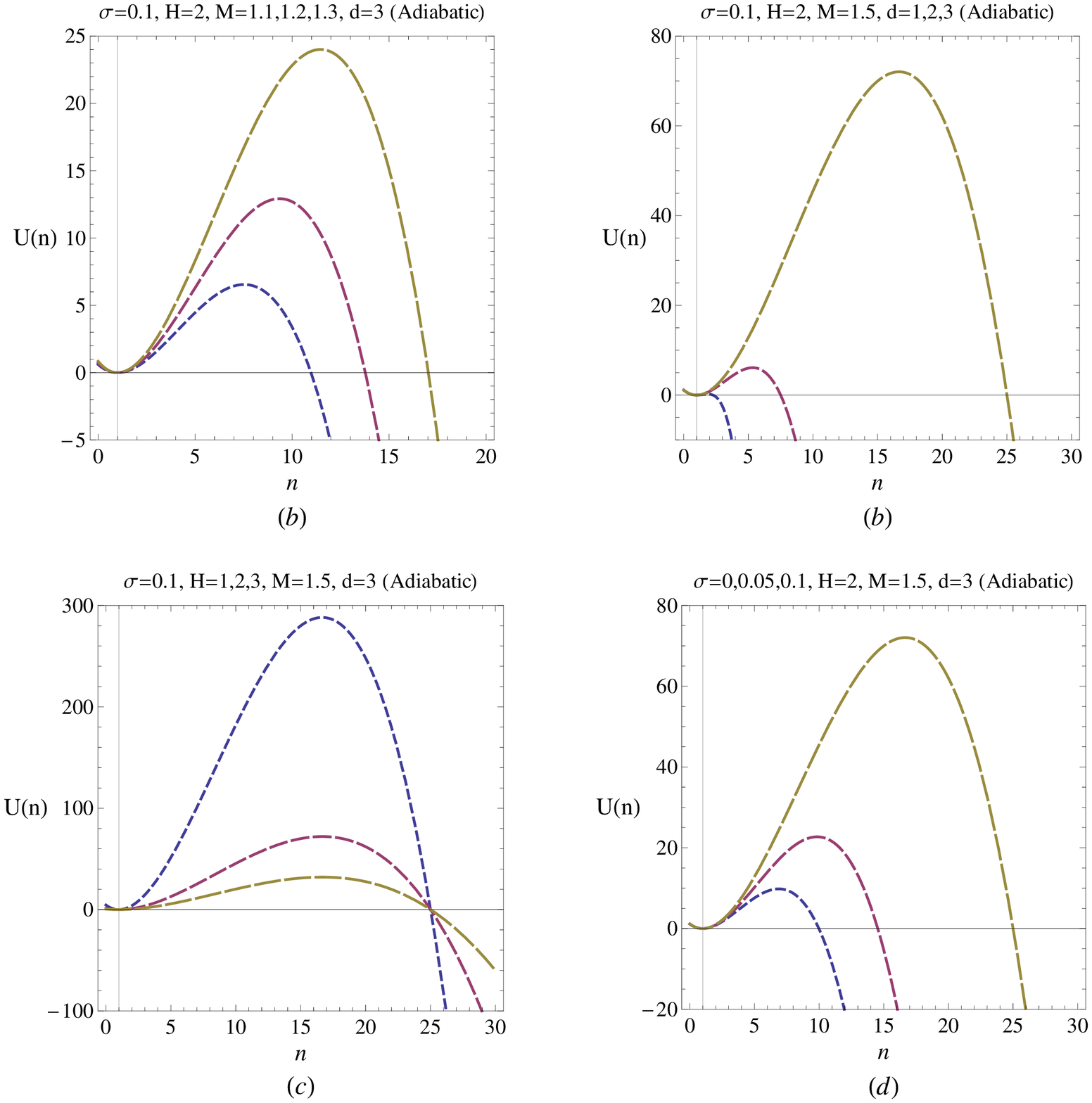}}
\caption{}
\label{fig:6}
\end{figure}

\end{document}